\documentclass[twocolumn]{IEEEtran}
\usepackage[normalem]{ulem} 
\usepackage{pgfplots}
\pgfplotsset{compat=newest}
\usetikzlibrary{plotmarks}
\usetikzlibrary{arrows.meta}
\usetikzlibrary{positioning}
\usepgfplotslibrary{patchplots}
\usepackage{grffile}
\usepackage{amsmath}
\usepackage{color, colortbl}
\definecolor{applegreen}{rgb}{0.7, 1, 0.0}
\definecolor{LightCyan}{rgb}{0.88,1,1}
\pgfplotsset{plot coordinates/math parser=false}
\newlength\figureheight
\newlength\figurewidth
  
\usepackage{breqn}
\usepackage{flafter} 
\usepackage{placeins}

\usepackage{siunitx}
\usepackage{graphicx}
\usepackage[english]{babel}
\usepackage{eqnarray}

\usepackage{multirow}
\usepackage[noadjust]{cite} 
\usepackage{arydshln}
\usepackage{longtable}
\usepackage{multirow}
\usepackage{array}

\usepackage{url}
\newcolumntype{M}[1]{>{\centering\arraybackslash}m{#1}}
\newcolumntype{N}{@{}m{0pt}@{}}

\usepackage{amssymb}
\usepackage{color}
\usepackage{textcomp}
\usepackage{diagbox}

\usepackage{subfigure}
\usepackage[percent]{overpic}
\usepackage{float}
\usepackage{algorithm}
\usepackage{algpseudocode}

\makeatletter
\renewcommand{\ALG@beginalgorithmic}
\makeatother

\errorcontextlines\maxdimen
\makeatletter
\newcommand*{\algrule}[1][\algorithmicindent]{\makebox[#1][l]{\hspace*{.5em}\thealgruleextra\vrule height \thealgruleheight depth \thealgruledepth}}%
\newcommand*{\thealgruleextra}{}
\newcommand*{\thealgruleheight}{1\baselineskip}
\newcommand*{\thealgruledepth}{.3\baselineskip}

\newcount\ALG@printindent@tempcnta
\def\ALG@printindent{%
    \ifnum \theALG@nested>0
        \ifx\ALG@text\ALG@x@notext
        \else
            \unskip
            \addvspace{-1pt}
            \ALG@printindent@tempcnta=1
            \loop
                \algrule[\csname ALG@ind@\the\ALG@printindent@tempcnta\endcsname]%
                \advance \ALG@printindent@tempcnta 1
            \ifnum \ALG@printindent@tempcnta<\numexpr\theALG@nested+1\relax
            \repeat
        \fi
    \fi
    }%
    
    \usepackage{etoolbox}
\patchcmd{\ALG@doentity}{\noindent\hskip\ALG@tlm}{\ALG@printindent}{}{\errmessage{failed to patch}}
\makeatother

\newbox\statebox
\newcommand{\myState}[1]{%
    \setbox\statebox=\vbox{#1}%
    \edef\thealgruleheight{\dimexpr \the\ht\statebox+1pt\relax}%
    \edef\thealgruledepth{\dimexpr \the\dp\statebox+1pt\relax}%
    \ifdim\thealgruleheight<.75\baselineskip
        \def\thealgruleheight{\dimexpr .75\baselineskip+1pt\relax}%
    \fi
    \ifdim\thealgruledepth<.25\baselineskip
        \def\thealgruledepth{\dimexpr .25\baselineskip+1pt\relax}%
    \fi
    \State #1%
    \def\thealgruleheight{\dimexpr .75\baselineskip+1pt\relax}%
    \def\thealgruledepth{\dimexpr .25\baselineskip+1pt\relax}%
}
\newcommand\longvdots[1]{\raisebox{1em}{\rotatebox{-90}{\hbox to #1 {\dotfill}}}}              
\usepackage{bbding} 

\usepackage{graphicx}
\usepackage[english]{babel}
\usepackage{eqnarray,amsmath}
\usepackage{multirow}
\usepackage[noadjust]{cite}
\usepackage[americanvoltages,fulldiodes]{circuitikz}
\usepackage{circuitikz}
\usepackage{siunitx}
\usepackage{tikz}
\usetikzlibrary{shapes,positioning}

\usepackage{amssymb}
\usepackage{color}
\usepackage{textcomp}
\usepackage{comment}

\usepackage{hyperref}
\hypersetup{
    colorlinks=true,
    linkcolor=blue,
    filecolor=magenta,      
    urlcolor=blue,
}

\newcommand{\ihab}[1]{{\textcolor{black}{#1}}}

\usepackage{eso-pic}
\newcommand\AtPageUpperMyright[1]{\AtPageUpperLeft{%
 \put(\LenToUnit{0.5\paperwidth},\LenToUnit{-1.5cm}){%
     \parbox{0.5\textwidth}{\raggedright\fontsize{11}{11}\selectfont #1}}%
 }}%
\newcommand{\conf}[1]{%
\AddToShipoutPictureBG*{%
\AtPageUpperMyright{#1}
}
}

\begin{document}
\title{Artificial Neural Network-Based Voltage Control of DC/DC Converter for DC Microgrid Applications}
\conf{\hspace*{-9cm}\textcolor{gray}{This paper has been accepted for publication at the 6th IEEE Workshop on the Electronic Grid} \textcolor{blue}{\href{https://ieee-egrid.org/}{(eGrid 2021)}}}
\author{\IEEEauthorblockN{Hussain Sarwar Khan\IEEEauthorrefmark{1}\Envelope, 
Ihab S. Mohamed\IEEEauthorrefmark{2}\Envelope, Kimmo Kauhaniemi\IEEEauthorrefmark{1}, and Lantao Liu\IEEEauthorrefmark{2} 
}\\
\IEEEauthorblockA{\IEEEauthorrefmark{1}School of Technology and Innovations, University of Vaasa, Vaasa, Finland
}\\
\IEEEauthorblockA{\IEEEauthorrefmark{2}Luddy School of Informatics, Computing, and Engineering, Indiana University, Bloomington, IN 47408, USA
} \\
\thanks{\textbf{e-mails \! \! in-order}:  {\href{hussain.khan@uwasa.fi}{\texttt{hussain.khan@uwasa.fi}}, \href{mohamedi@iu.edu}{\tt \texttt{mohamedi@iu.edu}}, \href{kimmo.kauhaniemi@uwasa.fi}{\texttt{kimmo.kauhaniemi@uwasa.fi}},
\href{lantao@iu.edu}{\texttt{lantao@iu.edu}}}\\
\Envelope \;Corresponding authors: Hussain Sarwar Khan and Ihab S. Mohamed} 
}

\maketitle

\begin {abstract}
The rapid growth of renewable energy technology enables the concept of microgrid (MG) to be widely accepted in the power systems. Due to the advantages of the DC distribution system such as easy integration of energy storage and less system loss, DC MG attracts significant attention nowadays. The linear controller such as PI or PID is matured and extensively used by the power electronics industry, but their performance is not optimal as system parameters are changed. In this study, an artificial neural network (ANN) based voltage control strategy is proposed for the DC--DC boost converter.  In this paper, the model predictive control (MPC) is used as an expert, which provides the data to train the proposed ANN. As ANN is tuned finely, then it is utilized directly to control the step-up DC converter. The main advantage of the ANN is that the neural network system identification decreases the inaccuracy of the system model even with inaccurate parameters and has less computational burden compared to MPC due to its parallel structure. To validate the performance of the proposed ANN, extensive MATLAB/Simulink simulations are carried out. The simulation results show that the ANN\ihab{-based control strategy} has better performance under different loading conditions \ihab{comparison to the PI controller}. The accuracy of the trained ANN model is \ihab{about} 97{\%}, which makes it suitable to be used for DC microgrid applications.

\end {abstract}


\section{Introduction}\label{Introduction}
\begin{figure}[!ht]
\renewcommand{\figurename}{Fig.}
\begin{center}
\input{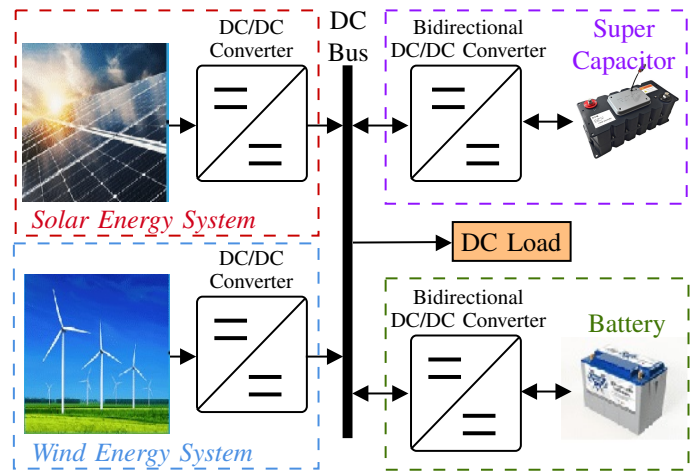}
\caption{Possible structure of single bus DC MG, \ihab{including renewable energy sources such as solar and wind, energy storage systems, and DC load}.}
\label{fig:DCMG}
\end{center}
\end{figure}

Renewable energy got attention due to the depletion of fossil fuels and global warming. Due to this, the use of DC/DC converters is rapidly increasing in a vast amount of applications such as wind turbines, photovoltaic systems, electric vehicles, energy storage systems, and in such applications, where different voltage levels loads are connected \cite{khan2021improved, khan2019finite,sahu2004low}. The block diagram of DC MG is expressed in Fig.~\ref{fig:DCMG}. DC MG mainly includes renewable energy sources (RES) such as solar \ihab{and} wind, energy storage system (\textit{ESS}), \ihab{and} DC load. Every RES and ESS is connected with the bus through a power electronic interface (PEI). It is therefore necessary to have effective control for the PEIs. The research community extensively proposes different types of linear controllers such as \ihab{proportional–integral} (PI) and \ihab{proportional–integral–derivative} (PID) and is vastly used by the PE industry \cite{hwu2009performance, hsu2009self, mohamed2013classical}. However, the linear controller has its practical limitations such as tuning of gains, poor disturbance rejection capability, shifting of the operating point of the converter towards instability due to change of the system parameters, and lacking the capacity to handle the non-linearities of the power system.
Many nonlinear control techniques such as model predictive control (\textit{MPC}), sliding mode control (\textit{SMC}), fuzzy-logic control (\textit{FLC}) have been proposed to cope with the issues mentioned above and also try to improve the transient behavior. In \cite{wu2000fuzzy}, FLC for DC converter is presented for PV-based lighting systems. FLC implementation for DC power converter using microcontroller has been studied in \cite{gupta1997implementation}. FLC basically works on the if-else statement, and its response depends upon predefined rules using if-else logic. FLC does not need any mathematical system model and also has the ability to handle the non-linearity of the system. Voltage regulation of FLC for DC/DC converter is also good under different conditions. However, many studies prove it as an unreliable controller because it lacks formal analysis. So, the amalgam of varying control techniques is found in the literature to balance the disadvantages of FLC \cite{jawhar2006neuro}.
\begin{figure*}[!ht]
\renewcommand{\figurename}{Fig.}
\begin{center}
\includegraphics[width=7.2in]{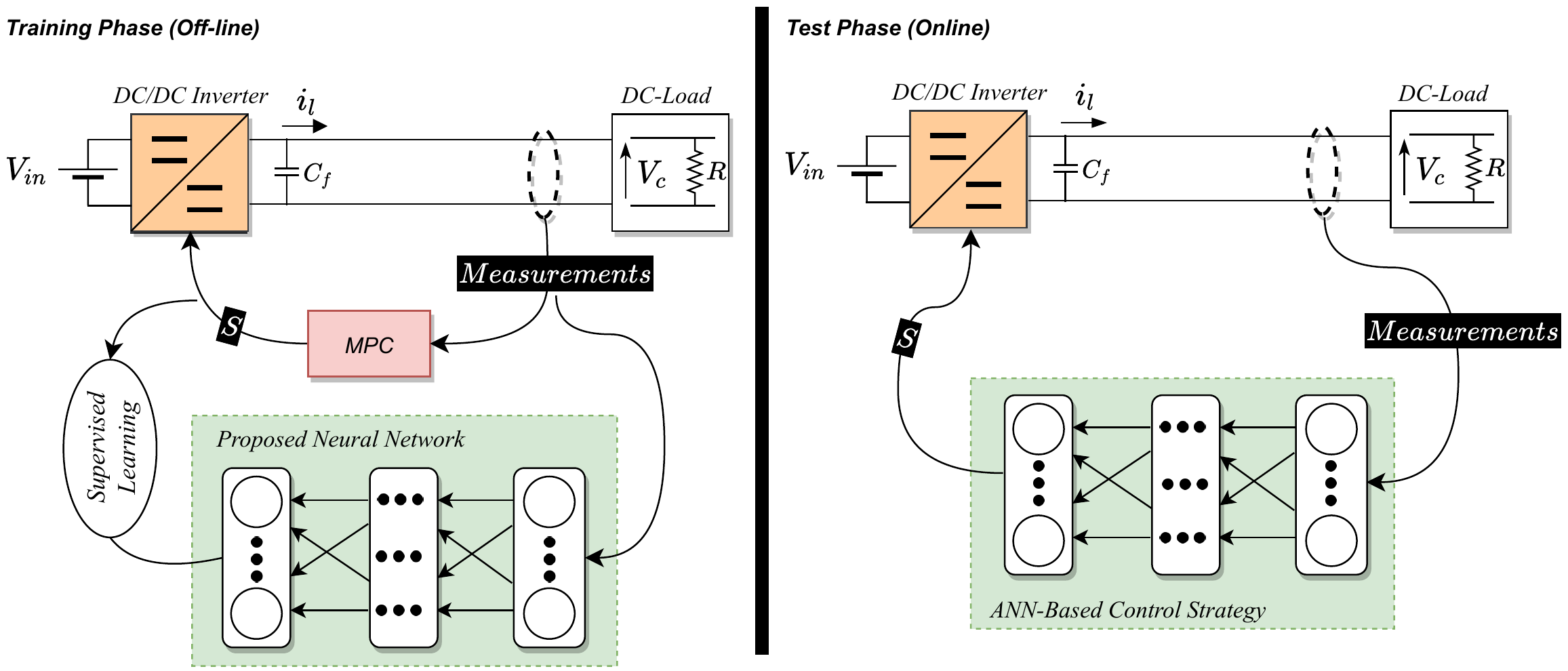}
\caption{Overview of the proposed control strategy. During the training phase, the classical MPC is used to control the DC/DC converter and collect the training data. In the test phase, the trained ANN is implemented to control the voltage of the converter instead of MPC  \cite{mohamed2019neural}.}
\label{fig:over}
\end{center}
\end{figure*}

Slide mode control and model predictive control  are developed and vastly studied in literature and have \ihab{become} promising solutions for power electronic converter applications. Slide mode control is based on variable structure control theory. Its basic principle is divided into two stages. The system \ihab{state} trajectory is forcefully taken into the user-defined sliding layer. This phase is known as the reaching phase, then in the second phase, which is known as the sliding phase, state trajectories remain within the layer, \ihab{defined} by the user on the base of application. 
It has better performance, robust against parametric variations, and possesses magnificent transient response under different loading conditions. Still, chattering phenomena, high switching losses, and complex mathematical modeling are the main barriers to its implementation \cite{oucheriah2012pwm, khan2021fault}.

MPC is a digital control method, and its basic principle is different from linear control. It uses the discrete-time model of the converter along with its filter to anticipate the behavior for all possible input combinations. One of the inputs having the least \ihab{(i.e., optimal)} value of the predefined cost function (\textit{CF}) is selected and applied to the coming sampling instant despite drafting a separate loop for each controlled variable and cascading them together as in the case of linear controllers \cite{mohamed2015improved}. CF is basically a square of the Euclidean distance between controlled and reference signals. However, it has a high computational burden, and its performance depends upon the mathematical model of the system and also has variable switching frequency; however, many new studies proposed a constant switching frequency based MPC for different power electronic applications \cite{karamanakos2013direct, cheng2017model,khan2021fault}.

Data-driven or model-free control techniques and especially ANN-based methods are growing in the domain of power converters \cite{lin1995power}. \ihab{An} ANN-based control \ihab{scheme} has been proposed in \cite{mohamed2019neural} \ihab{to directly control a three-phase inverter with an output LC filter, where a lower THD and a better steady and dynamic performance are achieved. Similarly, authors in \cite{malidarreh2021artificial} proposed an ANN-based control strategy for a three-phase flying capacitor multi-level inverter (FCMLI).} In \cite{saadatmand2020voltage}, \ihab{a} neural network predictive-based voltage control is proposed for the DC/DC buck converter. The author used PID controller data to train NN. After training, neural network predictive control (NNPC) is used to regulate the voltage. NNPC controller for grid-connected synchronverter is proposed in \cite{saadatmand2019neural}. 

\ihab{Broadly speaking, the} ANN\ihab{-based} controllers \ihab{are} better as compared to other controllers due to the following reasons \cite{mohamed2019neural,zhao2020overview}:
\begin{itemize}
\item They \ihab{do not} require \ihab{an explicit}
 mathematical model of the system.
\item Their performance is better if \ihab{they} are finely tuned \ihab{with sufficient data and properly chosen input features \cite{malidarreh2021artificial}}. 
\item \ihab{They} can be designed without having expert knowledge.
\end{itemize}

This paper proposes an artificial neural network-based voltage control for a DC/DC step-up (i.e., Boost) converter for DC microgrid applications. Initially, MPC-based voltage control is implemented for \ihab{the Boost} converter to extract the input features data. After the extraction of required data, a different possible combinations of inputs features are chosen. Finally, the voltage reference, inductor current, and capacitor voltage are selected as input features, while the converter switching \ihab{state} is taken as output feature for \ihab{the proposed} ANN in this study\ihab{, as illustrated in Fig.~\ref{fig:ann}}. Then, these combinations are used to train the ANN. Once the ANN is trained and has good model accuracy, the ANN model is \ihab{directly} used to generate the optimal switching \ihab{state} for the DC converter. Figure~\ref{fig:over} illustrates the overview of the proposed control strategy: the training phase combines using MPC to anticipate the converter output voltage converter and collection of state variables data under full-state observation. The collected data is used to train the ANN. In the test phase, the trained neural network is employed online to control the converter's output voltage instead of MPC. The simulation results of the proposed control strategy are also compared with the traditional PI Controller.

The rest of the paper is organized as follows. \ihab{The} mathematical modeling of the DC/DC boost converter and the basic principle of MPC \ihab{are} explained in Section \ref{mathematical-Modeling}. \ihab{While the proposed} ANN and \ihab{its} training \ihab{procedure} are elaborated in Section \ref{ANN}. Section \ref{simulation} shows the simulation results \ihab{for both ANN and PI controllers}. Then, future work is discussed in Section \ref{fwork}. Finally, Section \ref{CONCLUSION} presents the conclusion.

\begin{figure}[!ht]
\renewcommand{\figurename}{Fig.}
\centering
\includegraphics[width=0.95\columnwidth]{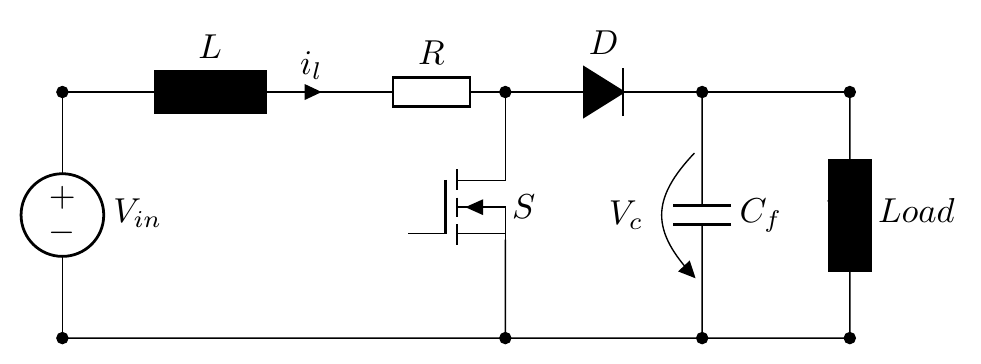}
\caption{Circuit diagram of the DC/DC converter.}
\label{fig:boostckt}
\end{figure}

\section{Mathematical Modelling of MPC}\label{mathematical-Modeling}
\ihab{Figure~\ref{fig:boostckt} illustrates the circuit diagram of the simple step-up boost converter, where $S$ is a controllable switch, $R$ is the damping resistance, the current through the inductor $L$ is $i_{l}$, and the voltage across the capacitor $C_{f}$ is $V_{c}$. $V_{in}$ represents the input voltage of the DC source. The second order low pass filter is used to attenuate the ripples and distortion.}
In order to implement the MPC, it is necessary to develop the discrete time model of the DC converter. The output voltage of the step-up DC converter, \ihab{namely, $V_c$}, is controlled by varying the duty cycle of pulse width modulation (PWM) signal. But in the case of MPC,  \ihab{the} pulse for the switch is directly generated by the MPC. One of the major drawbacks of MPC is its variable switching frequency. 
Equation (\ref{eq:1}) explains the inductive nature, while (\ref{eq:2}) presents the capacitive behavior of the system:
\begin{equation}
\frac{d i_{l}(t)}{d(t)}=-\frac{R} {L} i_{l}(t)-\frac{V_{c}(t)}{L}+\frac{V_{c}(t)}{L} u(t)+\frac{V_{\text {in }}}{L},
\label{eq:1}
\end{equation}
\begin{equation}
\frac{d V_{c}(t)}{d(t)}=\frac{1}{C_{f}} i_{L}(t)-\frac{1}{C_{f}} i_{L}(t) u(t)-\frac{1}{R C_{f}} V_{c}(t).
\label{eq:2}
\end{equation}
The switch states, $S$, are defined by function $u(t)$, as shown in (\ref{eq:3}). If $S=1$, then switch $S$ is in ON state and if switch $S=0$, then switch $S$ is in OFF state.
\begin{equation}
u(t)= \begin{cases}1, & \text{if} \; S=1 \\ 0, & \text{if} \; S=0\end{cases}
\label{eq:3}
\end{equation}
The discrete-time model of the DC converter is expressed in (\ref{eq:4}) and (\ref{eq:5}). These equations are used to anticipate the future response of voltage and current.
\begin{equation}
i_{l}(k+1)=\left(\frac{T R}{L}-1\right) i_{l}(k)+(u(k)-1) \frac{T}{L} V_{c}(k)
\label{eq:4}
\end{equation}
\begin{equation}
\begin{aligned}
&V_{c}(k+1)=\frac{T}{C_{f}} i_{l}(k)+\left(1-\frac{T}{C_{f} R}\right) V_{c}(k)-
\frac{T}{C_{f}} i_{l}(k){u}(k)
\end{aligned}
\label{eq:5}
\end{equation}
Where $k+1$ represents the future or next (coming) instant and $T$ is the sampling time.

The formulation of cost function (\textit{CF}) is an essential part of the development of MPC, and it is the positive value of error between the reference and actual value of the state parameter. The CF, $J$, chosen in this study is illustrated in (\ref{eq:6}), \ihab{where $V_{c}^{*}$ is the reference  output voltage}. 
\begin{equation}
J(k)=\left(V_{c}^{*}(k+1)^{2}-V_{c}(k+1)\right)^{2}
\label{eq:6}
\end{equation}
The execution of the MPC algorithm can be summarized as follows:
\begin{itemize}
\item At the start of the switching instant, the voltage and current of the converter are measured using sensors.
\item Equations (\ref{eq:4}) and (\ref{eq:5}) are used to predict the current and voltage at \ihab{instant $k+1$} for all possible switching states, and then the CF is evaluated using (\ref{eq:6}) for all possible states. In this study, $N$ is taken as one. So, there is only two possible switching states.
\item The switching state that minimizes the CF is applied to the converter at the next time instant $k+1$.
\end{itemize}
\section{Proposed ANN-Based Control Strategy}\label{ANN}
Basically, ANN is a network \ihab{that has} one or more hidden layers and each layer has one or multiple neurons which makes the ANN response similar to the real neural network. In this study, \ihab{the} feed-forward \ihab{ANN} is used\ihab{, which is called FF-ANN}. In FF-ANN, the data moves in forward direction only. The output of single neuron is mathematical expressed as:
\begin{equation}
y =\operatorname{Act}\left(b+\sum_{i=1}^{M} x_{i}w_{i}\right), 
\label{eq:7}
\end{equation}
where $Act(.)$, $w_{i}$, $b$, and $M$ are the activation function, weights of each input $x_i$, bias or correction factor, and number of input elements (or neurons) where the input features $x = \{x_1, x_2, \dots, x_M\}$, respectively. \ihab{The most commonly used types of} activation functions are given in Table \ref{table:1}. By joining the multiple neurons into a single layer, \ihab{an} FF-ANN layer can be developed. The general equation used to compute the output of the \ihab{multi-input single-output} FF-ANN \ihab{can be expressed} as:
\begin{equation}
\begin{aligned}
y_1&=Act\left(\sum_{j=1}^{J} {}^2w_{j1} h_{j}+{}^2b_{1}\right), \text{and}\\
h_{j}&=Act\left(\sum_{m=1}^{M} {}^1w_{mj} x_{m}+{}^1b_{j}\right), \quad \forall j = \{1, \dots, J\},
\end{aligned}
\label{eq:8}
\end{equation}
\ihab{where $y_{1}$ is the output of the ANN, $({}^1w_{mj},{}^2w_{j1})$ represent the weights of the hidden and output layers, $J$ represents the number of hidden layers, $M$ represents the number of input neurons, and $({}^1b_{j}, {}^2b_{1})$ refer to the biases of the hidden and output layers, respectively.} 
\begin{figure}[!ht]
\renewcommand{\figurename}{Fig.}
\begin{center}
\hspace*{-3pt}
 
 

\tikzset{%
every input neuron/.style={circle, draw, fill=green!50, text width = 0.2cm, inner sep = 0.2cm},
every output neuron/.style={circle, draw, fill=red!50, text width = 0.2cm, inner sep = 0.2cm},
every hidden neuron/.style={circle, draw, fill=blue!50, text width = 0.2cm, inner sep = 0.2cm},
neuron missing/.style={ draw=none, scale=3, fill=none, text height=0.333cm, execute at begin node=\color{black}$\vdots$},
neuron2 missing/.style={ draw=none, scale=3, fill=none, text height=0.15cm, execute at begin node=\color{black}$\vdots$},
}

\begin{tikzpicture}[x=1.2cm, y=1.3cm, >=latex, line width=0.25mm]
\foreach \m/\l [count=\y] in {1,2,3}
  \node [every input neuron/.try, neuron \m/.try] (input-\m) at (0,1.6-1*\y) {};
\foreach \m [count=\y] in {1,missing,2}
  \node [every hidden neuron/.try, neuron2 \m/.try ] (hidden-\m) at (2,2.-\y*1.2) {};
\foreach \m [count=\y] in {1}
  \node [every output neuron/.try, neuron2 \m/.try ] (output-\m) at (3.8,0.7-\y*1.) {};

\foreach \l [count=\i] in {1}
  \node at (input-\l) {$x_\l$};
  \draw [<-] (input-1) -- ++(-0.9,0)
    node [above] {$V^*_c$};
    
\foreach \l [count=\i] in {2}
    \node at (input-\l) {$x_\l$};
  \draw [<-] (input-2) -- ++(-0.9,0)
    node [above] {$V_c$};
    
\foreach \l [count=\i] in {3}
  \node at (input-\l) {$x_\l$};
  \draw [<-] (input-3) -- ++(-0.9,0)
    node [above] {$i_l$};

\foreach \l [count=\i] in {1,J}
  {
    \node at (hidden-\i) {$h_\l$};
    \draw [latex-, rotate=90] (hidden-\i) -- ++(0.6,0)
    node [above] {${}^1b_{\l}$};
  }
  
\foreach \l [count=\i] in {1}
  {
  \node at (output-\i) {$y_\i$};
  \draw [-latex] (output-\i) -- ++(1.1,0)
    node [above,midway] {$S_{opt}$};
    \draw [latex-, rotate=90] (output-\i) -- ++(0.6,0)
    node [above] {${}^2b_{\l}$};
  }
  
\foreach \i in {1,...,3}
  \foreach \j in {1,...,2}
    \draw [->] (input-\i) -- (hidden-\j);
    
\foreach \i in {1}
  \foreach \j [count=\c] in {1,J}
    \draw [-latex] (input-\i) --  node [pos=0.6,fill=cyan!10,inner sep=1pt, rounded corners]{${}^1w_{\i\j}$} (hidden-\c);
\foreach \i in {3}
  \foreach \j [count=\c] in {1,J}
    \draw [-latex] (input-\i) --  node [pos=0.6,fill=cyan!10,inner sep=1pt, rounded corners]{${}^1w_{3\j}$} (hidden-\c);
    
\foreach \i in {1}
  \foreach \j in {1}
    \draw [->] (hidden-\i) -- (output-\j);
    
\foreach \i in {1}
  \foreach \j [count=\c] in {1}
    \draw [-latex] (hidden-\i) --  node [pos=0.4,fill=cyan!10,inner sep=1pt, rounded corners]{${}^2w_{\i\j}$} (output-\c);
    
\foreach \i in {2}
  \foreach \j [count=\c] in {1}
    \draw [-latex] (hidden-\i) --  node [pos=0.5,fill=cyan!10,inner sep=1pt, rounded corners]{${}^2w_{J\j}$} (output-\c);
  
  
\draw (0.1,2) node [anchor=north west][inner sep=0.75pt]   [align=left] {\small \textcolor{red}{\textit{ANN}-Based Control Strategy}};
\node[rectangle,
    draw = red,
    dash pattern={on 3.75pt off 3pt on 7.5pt off 1.5pt},
    minimum width = 5.8cm, 
    minimum height = 5.28cm] (r) at (1.8,0.05) {};
 \node[rectangle,
    draw = blue,
    minimum width = 1.7cm,
    line width=0.4mm,
    minimum height = 1.1cm] (r) at (5.59,-0.3) {
    \begin{minipage}[lt]{38pt}\setlength\topsep{0pt}
        \begin{center}
        \textcolor{blue}{DC/DC}  
        \end{center}
        \textcolor{blue}{Converter}
        \end{minipage}};
\end{tikzpicture}
\caption{Block diagram of the proposed  ANN-based control scheme for the DC/DC converter. It is trained to map directly from the measured variables, namely, $V^*_c$, $V_c$, and $i_l$, to the \textit{optimum}  switching state $S_{opt}$.}
\label{fig:ann}
\end{center}
\end{figure}
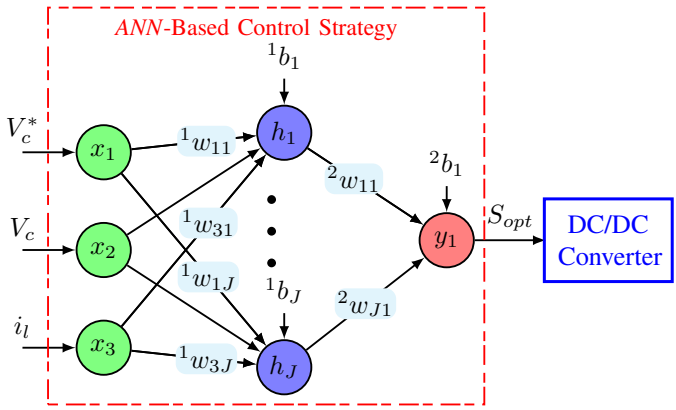
\begin{table}[!ht]
\small\addtolength{\tabcolsep}{0.0pt} 
\setlength\extrarowheight{5pt}
\caption{Activation function types.}
\centering
\begin{tabular}{|l || c|} 
\hline 
\rowcolor{LightCyan}
 $Act(.) \equiv f(x)$ & Definition \\
 \hline\hline  
 \textit{Sigmoid} & $ f(x) = \frac{1}{1+e^{-x}}$\\
 \hline
\textit{Hyperbolic Tangent (tanh)}  & $f(x) =  \frac{e^{x}-e^{-x}}{e^{x}+e^{-x}}$\\
 \hline
 \textit{Rectified Linear} & $f(x)= \begin{cases}0 & x \leq 0 \\ x & x>0\end{cases}$\\
 \hline
 \textit{Binary Step} & $f(x)=\left\{\begin{array}{ll}
0 & x<0 \\
1 & x>0
\end{array}\right.$\\
\hline
\end{tabular}
\label{table:1}
\end{table}

Figure \ref{fig:ann} demonstrates the proposed ANN-based control strategy used in this study. The accuracy of MPC depends on the mathematical modeling of the system. However, the proposed control scheme does not need the model of the system, but it requires the training dataset. It maps \ihab{directly from the raw input features to the desired outputs}. Therefore, the performance of the ANN does not depend upon the system model or its parameter. 
\ihab{In this work,} the reference voltage $V^*_c$, capacitor voltage  $V_c$ and inductor current $i_l$ are chosen as the input features of the trained ANN-based control strategy, while the optimal switching state $S_{opt}$ is considered as its target or output. 
Initially, the MPC algorithm is simulated to extract the training \ihab{data, which consists of the input features and the corresponding output, i.e., input-output pairs}.
Then, the extracted data is used to train the ANN. In our case, the total number of training data samples is $30001$. The control-loop of the proposed ANN-based control strategy, at instant $k$ is summarized as follows:
\begin{enumerate}
\item  Initially, measure $i_{l}$ and $V_{c}$ at instant $k$.
\item  Those measured variables, along with the reference value $V^*_{c}$, are utilized by our proposed controller to directly predict the optimal switching state $S_{opt}$.
\item  Then, the optimal switching state is directly applied to the converter without using any modulator. 
\end{enumerate}
\begin{figure}[!ht]
\renewcommand{\figurename}{Fig.}
\begin{center}
%
%
\definecolor{mycolor1}{rgb}{0.73725,0.90196,0.76863}%
\definecolor{mycolor2}{rgb}{0.97647,0.76863,0.75294}%
\definecolor{mycolor3}{rgb}{0.94118,0.94118,0.94118}%
\definecolor{mycolor4}{rgb}{0.13333,0.67451,0.23529}%
\definecolor{mycolor5}{rgb}{0.88627,0.23922,0.17647}%
\definecolor{mycolor6}{rgb}{0.85098,0.85098,0.85098}%
\begin{tikzpicture}[scale=1]

\begin{axis}[%
width=2in,
height=2in,
at={(0.545in,0.545in)},
scale only axis,
xmin=0.5,
xmax=3.5,
xtick={1,2,3,4},
xticklabels={{0},{1},{}},
xticklabel style={rotate=45},
xlabel style={font=\bfseries\color{white!15!black}},
xlabel={Target Class},
y dir=reverse,
ymin=0.5,
ymax=3.5,
ytick={1,2,3,4},
yticklabels={{0},{1},{}},
ylabel style={font=\bfseries\color{white!15!black}},
ylabel={Output Class},
axis background/.style={fill=white},
title style={font=\bfseries},
legend style={legend cell align=left, align=left, draw=white!15!black}
]

\addplot[area legend, draw=black, fill=mycolor1]
table[row sep=crcr] {%
x	y\\
0.5	0.5\\
1.5	0.5\\
1.5	1.5\\
0.5	1.5\\
}--cycle;

\node[above, align=center, font=\bfseries]
at (axis cs:1,1) {6342};
\node[below, align=center]
at (axis cs:1,1) {21.1\%};

\addplot[area legend, draw=black, fill=mycolor2]
table[row sep=crcr] {%
x	y\\
0.5	1.5\\
1.5	1.5\\
1.5	2.5\\
0.5	2.5\\
}--cycle;

\node[above, align=center, font=\bfseries]
at (axis cs:1,2) {825};
\node[below, align=center]
at (axis cs:1,2) {2.7\%};

\addplot[area legend, draw=black, fill=mycolor3]
table[row sep=crcr] {%
x	y\\
0.5	2.5\\
1.5	2.5\\
1.5	3.5\\
0.5	3.5\\
}--cycle;

\node[above, align=center, font=\color{mycolor4}]
at (axis cs:1,3) {88.5\%};
\node[below, align=center, font=\color{mycolor5}]
at (axis cs:1,3) {11.5\%};

\addplot[area legend, draw=black, fill=mycolor2]
table[row sep=crcr] {%
x	y\\
1.5	0.5\\
2.5	0.5\\
2.5	1.5\\
1.5	1.5\\
}--cycle;

\node[above, align=center, font=\bfseries]
at (axis cs:2,1) {0};
\node[below, align=center]
at (axis cs:2,1) {0.0\%};

\addplot[area legend, draw=black, fill=mycolor1]
table[row sep=crcr] {%
x	y\\
1.5	1.5\\
2.5	1.5\\
2.5	2.5\\
1.5	2.5\\
}--cycle;

\node[above, align=center, font=\bfseries]
at (axis cs:2,2) {22834};
\node[below, align=center]
at (axis cs:2,2) {76.1\%};

\addplot[area legend, draw=black, fill=mycolor3]
table[row sep=crcr] {%
x	y\\
1.5	2.5\\
2.5	2.5\\
2.5	3.5\\
1.5	3.5\\
}--cycle;

\node[above, align=center, font=\color{mycolor4}]
at (axis cs:2,3) {100\%};
\node[below, align=center, font=\color{mycolor5}]
at (axis cs:2,3) {0.0\%};

\addplot[area legend, draw=black, fill=mycolor3]
table[row sep=crcr] {%
x	y\\
2.5	0.5\\
3.5	0.5\\
3.5	1.5\\
2.5	1.5\\
}--cycle;

\node[above, align=center, font=\color{mycolor4}]
at (axis cs:3,1) {100\%};
\node[below, align=center, font=\color{mycolor5}]
at (axis cs:3,1) {0.0\%};

\addplot[area legend, draw=black, fill=mycolor3]
table[row sep=crcr] {%
x	y\\
2.5	1.5\\
3.5	1.5\\
3.5	2.5\\
2.5	2.5\\
}--cycle;

\node[above, align=center, font=\color{mycolor4}]
at (axis cs:3,2) {96.5\%};
\node[below, align=center, font=\color{mycolor5}]
at (axis cs:3,2) {3.5\%};

\addplot[area legend, draw=black, fill=mycolor6]
table[row sep=crcr] {%
x	y\\
2.5	2.5\\
3.5	2.5\\
3.5	3.5\\
2.5	3.5\\
}--cycle;

\node[above, align=center, font=\bfseries\color{mycolor4}]
at (axis cs:3,3) {97.3\%};
\node[below, align=center, font=\bfseries\color{mycolor5}]
at (axis cs:3,3) {2.7\%};
\addplot [color=darkgray, line width=2.0pt]
  table[row sep=crcr]{%
2.5	0.5\\
2.5	3.5\\
};

\addplot [color=darkgray, line width=2.0pt]
  table[row sep=crcr]{%
0.5	2.5\\
3.5	2.5\\
};

\end{axis}
\end{tikzpicture}%
\caption{Confusion matrix of the trained ANN based on the overall training data, where the correct and incorrect observations are highlighted in green and red, respectively.}
\label{fig:con}
\end{center}
\end{figure}
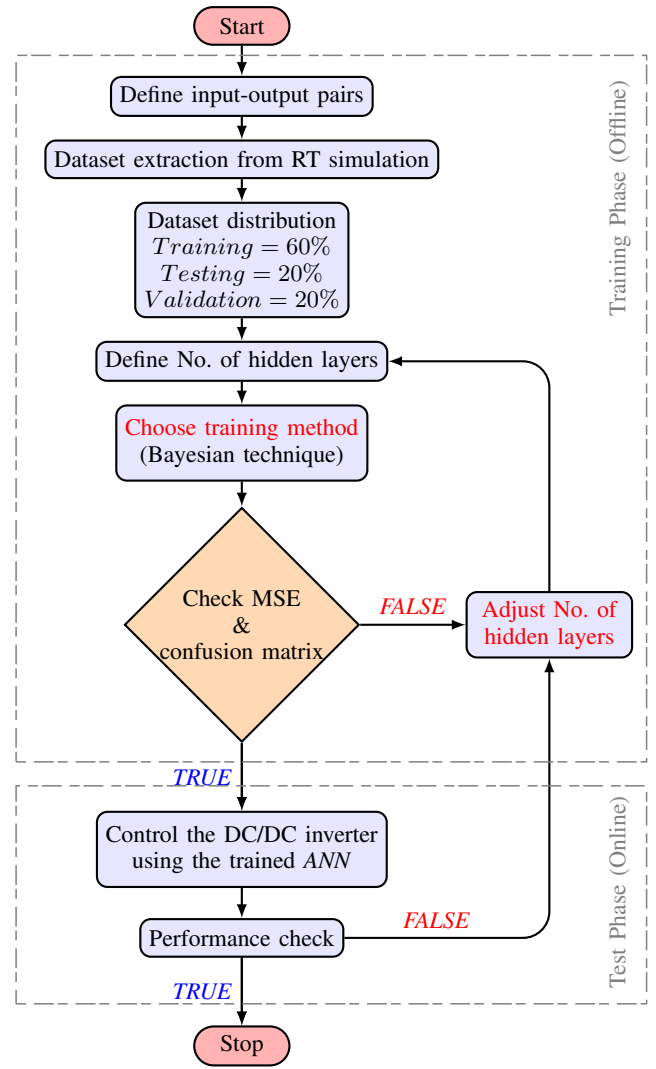
\begin{figure}[!ht]
\renewcommand{\figurename}{Fig.}
\centering
 
 

\hspace*{-1cm}\begin{tikzpicture}[font=\small,thick, scale=0.85]
 
\node[draw,
    rounded rectangle,
    minimum width=1.5cm,
    minimum height=0.5cm,
    fill=red!30] (block1) {Start};
    
\node[draw,
    rounded corners,
    below= 0.4cm of block1,
    minimum width=1.0cm,
    minimum height=0.5cm,
     fill=blue!10,
] (block2) {Define input-output pairs};

\node[draw,
    rounded corners,
    below= 0.3cm of block2,
    minimum width=1.0cm,
    minimum height=0.5cm,
     fill=blue!10,
] (block3) {Dataset extraction from RT simulation};

 \node[draw,
    rounded corners,
     align=center,
    below= 0.3cm of block3,
    minimum width=1.0cm,
    minimum height=1.4cm,
     fill=blue!10,
] (block4) {\textcolor{black}{Dataset distribution} \\
$ Training=60\%$  \\ 
$ Testing=20\%$ \\
$ Validation=20\%$  };

    
\node[draw,
    rounded corners,
    align=center,
    below = 0.3cm of block4,
    minimum width=3.3cm,
    minimum height=0.5cm,
    fill=blue!10,
    ] (block5) {Define No. of hidden layers};
    
\node[draw,
    align=center,
    rounded corners,
    below = 0.3cm of block5,
    minimum width=3.0cm,
    minimum height=1cm,
    fill=blue!10,
    ] (block6) {\textcolor{red}{Choose training method} \\ (Bayesian technique)
    };
    
 \node[draw,
    diamond,
    below= 0.33cm of block6,
    minimum width=1.6cm,
    fill=orange!30,
    inner sep=0.1,align=center] (block7) { Check MSE \\ \& \\ confusion matrix};
    
    \node[draw,
    align=center,
    rounded corners,
    below = 0.9cm of block7,
    minimum width=3.0cm,
    minimum height=1.cm,
    fill=blue!10,
    ] (block9) {Control the DC/DC inverter\\ using the trained \textit{ANN}};
    
\node[draw,
align=center,
    rounded corners,
    right = of block7,
    minimum width=2.2cm,
    minimum height=0.5cm,
    xshift=0.4cm,
    fill=blue!10,
    ] (block8){\textcolor{red}{Adjust No. of} \\ \textcolor{red}{hidden layers}};
 
\node[draw,
    rounded corners,
    below =0.4cm of block9,
    minimum width=1.0cm,
    minimum height=0.5cm,
    fill=blue!10,
    ] (block10) {Performance check};
    
\node[draw,
    rounded rectangle,
    below =0.9cm of block10,
    minimum width=1.5cm,
    minimum height=0.5cm,
    fill=red!30] (block11) {Stop};
    
\node [left=2.5 cm of block4] (com) {};
\coordinate [left=2.5cm of block4] (com);

\node [right= 3.8 cm of block4] (com2) {};
\coordinate [right=3.8 cm of block4] (com2);
    
\draw[-latex] (block1) edge (block2)
    (block2) edge (block3)
    (block3) edge (block4)
   (block4) edge (block5)
   (block5) edge (block6)
   (block6) edge (block7)
   (block7) edge (block9)
   (block9) edge (block10)
  (block10)edge (block11) ;
 \draw[-latex] (block7) edge node[anchor=south]{\textcolor{red}{\textit{FALSE}}} (block8);
 \draw[-latex, rounded corners=10pt] (block10) -| node[anchor=south] {\textcolor{red}{\hspace*{-3cm}\textit{FALSE}}} (block8);
 \draw[-latex, rounded corners=10pt] (block8) |- (block5);
\draw[-latex] (block7) edge node[anchor=east]{\textcolor{blue}{\textit{TRUE}}}(block9);
\draw[-latex] (block10) edge node[anchor=east]{\textcolor{blue}{\textit{TRUE}}}(block11);

\node[rectangle,
   draw = gray,
    dash pattern={on 3.75pt off 3pt on 7.5pt off 1.5pt},
    minimum width = 3.3in,
    line width=0.2mm,
    minimum height = 9.4cm] (l) at (1.4,-5.99) {};
    \draw (5.7,-0.8) node [anchor=north west][inner sep=0.75pt]   [align=left] {\textcolor{gray}{\rotatebox{90}{Training Phase (Offline)}}};
    
    \node[rectangle,
   draw = gray,
    dash pattern={on 3.75pt off 3pt on 7.5pt off 1.5pt},
    minimum width = 3.3in,
    line width=0.2mm,
    minimum height = 2.9cm] (l) at (1.4,-13.6) {};
    \draw (5.7,-12) node [anchor=north west][inner sep=0.75pt]   [align=left] {\textcolor{gray}{\rotatebox{90}{Test Phase (Online)}}};
\end{tikzpicture}
\caption{Main steps of deploying the ANN-based control strategy for the DC/DC converter.}
\label{fig:flow}
\end{figure}

A grid search tuning method is used for the selection of configuration with 15 neurons. Bayesian regularized technique (BRT) is used to train the ANN and adjust the biases and weights. BRT is more robust than standard propagation methods and can reduce or eliminate the need for lengthy cross-validation \cite{foresee1997gauss}. In this research work, 60$\%$ of the random input data is used to train the ANN, while 20$\%$ is used for testing and 20$\%$ validation. Figure~\ref{fig:con} presents the overall confusion matrix, which is used to analyze the accuracy of the trained ANN. The correct classification of the data class is presented in the diagonal entries of the matrix, while other entries show the incorrect classification of the data. The trained ANN that has been used, in this study, has an accuracy of 97$\%$. The trained ANN model is exported to Simulink to test its performance under the original scenario. 
\ihab{To sum up, the complete procedure of the learning-based control strategy is illustrated in Fig. \ref{fig:flow}, highlighting the main steps of the training and test phases.}

\section{Simulation Results}\label{simulation}
The trained ANN model is exported into the Simulink model of the DC/DC converter to validate and verify the performance of the proposed control strategy. Extensive MATLAB/Simulink simulation is carried out. The performance of the boost converter with the proposed control  scheme is investigated under normal load and step change of load. 
The simulation parameters of the converter are given in Table \ref{tab:2}.
\begin{table}[!ht]
\caption{Simulation Parameters.}
\centering
\begin{tabular}{l c} 
\hline
\rowcolor{LightCyan}
 Parameter & Value \\
 \hline\hline  
 DC Input $V_{in}$  & \SI{70}{\;[\volt]}\\
 Inductor value  $L$  &\SI{10e-3}{\;[\henry]}\\
 Resistance Value $R$ & \SI{80e-3}{\;[\Omega]}\\
 Capacitor $C_{f}$ & \SI{100}{\;[m\farad]}\\
 Load $P$ &  \si{0}$-$\SI{1500}{\;[\watt]}\\
 PI Parameter $K_p, K_i$ & \num{0.054}, \num{8.86} \\
 Switching Frequency & \SI{20}{\;[\kilo Hz]}\\
\hline
\end{tabular}
\label{tab:2}
\end{table}

\begin{figure}[!ht]
\renewcommand{\figurename}{Fig.}
\begin{center}
\includegraphics[width=1\columnwidth]{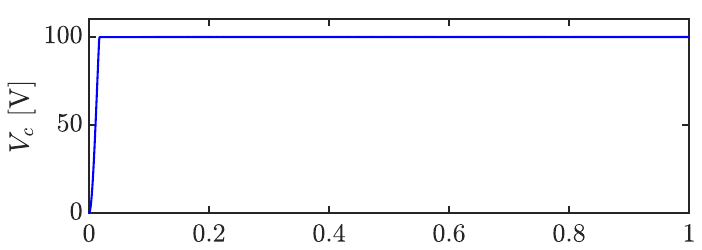}\\
\includegraphics[width=1\columnwidth]{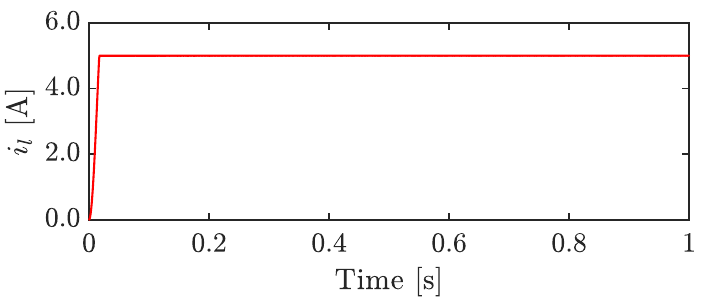}
\caption{Simulation results of the output voltage and current of the DC boost converter under normal load conditions.}
\label{fig:7}
\end{center}
\end{figure}

\begin{figure}[!ht]
\renewcommand{\figurename}{Fig.}
\begin{center}
\includegraphics[width=1\columnwidth]{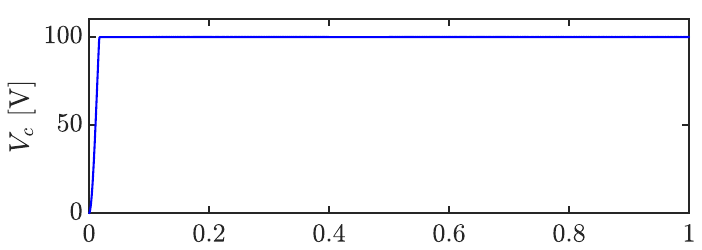}\\
\includegraphics[width=1\columnwidth]{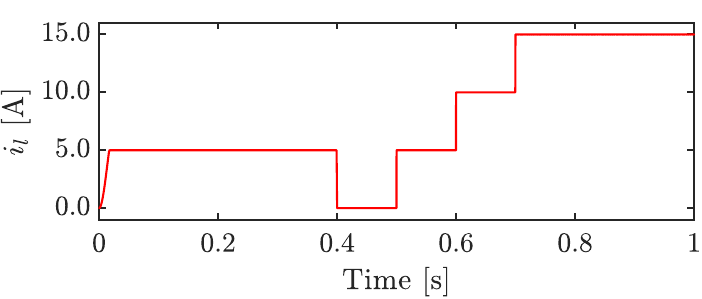}
\caption{Simulation results of the output voltage and current of the DC boost converter under full load to no-load test.}
\label{fig:8}
\end{center}
\end{figure}

Figure \ref{fig:7} illustrates the performance of our proposed control strategy, considering normal load conditions. The simulation starts at $t=\SI{0}{\second}$, where a resistive load of \SI{20}{\ohm} is connected with the system. The reference voltage is set to \SI{95}{\volt}. Initially, the system takes around \SI{20}{\milli\second} to reach the reference value. After a transient period, the output voltage and current wave forms remain stable and do not show distortion.
Figure \ref{fig:8} shows the performance of the proposed controller from full load to no-load condition and vice versa. At $t=\SI{0.4}{\second}$, the load is disconnected from the system; i.e., the converter is under no-load condition. It is observed that the voltage remains stable, and the current becomes zero. While at $t=\SI{0.5}{\second}$, the load is again connected to the system.
The voltage remains stable, while the current is increased to $\SI{4.9}{\ampere}$. However, there is no transient observed in the simulation. After the interval of \SI{0.6}{\second}, further loads are added into the system to investigate the response of the proposed controller. It is observed from Fig. \ref{fig:8} that with increasing the load, the voltage remains stable while keeping track of the reference value with un-noticeable distortion, demonstrating the superior performance of the proposed ANN-based control scheme under different loading and transient conditions. 
\begin{figure}[!ht]
\renewcommand{\figurename}{Fig.}
\begin{center}
\includegraphics[width=1\columnwidth]{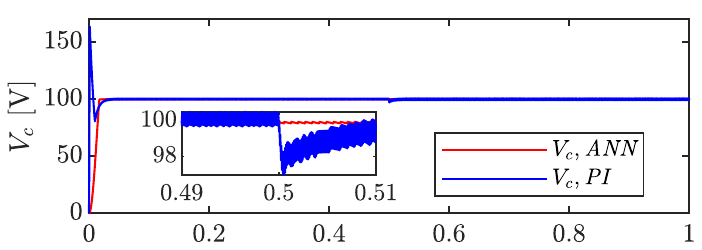}\\
\includegraphics[width=1\columnwidth]{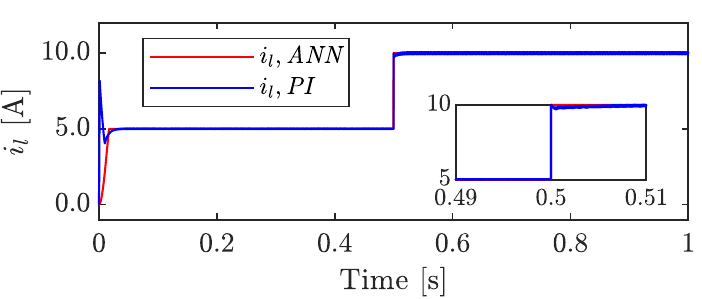}
\caption{Simulation results of the output voltage and current of the DC boost converter using PI and ANN-based controllers under step change of load.}
\label{fig:9}
\end{center}
\end{figure}
Figure \ref{fig:9} presents the simulation results of our proposed controller under step change of load. At $t=\SI{0.5}{\second}$, the DC load becomes double. We can observe that the voltage remains stable without any distortion, whereas the current increases with the increase of load. However, the current waveform becomes stable with almost no transient time. 
Figure \ref{fig:9} also presents a comparison with the PI controller. 
Under the transient period, the PI controller shows an overshoot in the voltage and current which may harm the switch of the DC/DC converter; accordingly, a high rating semi-conductor switch is required which increases the converter cost.
The ANN-based control scheme has better wave quality and less distortion compared to the PI controller. Moreover, the output current of the PI-based controlled converter is distorted, while the current wave in the case of ANN is constant, stable, and has less loss compared to the PI controller.

\section{Discussion and Future Work}\label{fwork}
In this study, our proposed control strategy is trained on a single reference value (i.e., $V^*_c = \SI{95}{\volt}$) and also tested on the same reference value under different loading conditions. In the future work, the proposed ANN model will be trained and its performance will be examined on different reference values. Moreover, it will be trained on various parameters such as filter values, switching frequency, etc, proposing a more generic control strategy.
We have also observed that the performance of our proposed controller is similar to the MPC used to extract the training data. For this reason, the comparison is carried out with the PI controller. However, the ANN-based controller has an advantage over MPC as it has less computation burden and constant switching frequency.

\section{Conclusion}\label{CONCLUSION}
Within this work, we proposed a feed-forward artificial neural network-based voltage control strategy for the DC/DC step-up converter. Model predictive control is implemented to extract the training data, where the data is used, later on, to train the ANN offline. 
After training the ANN properly, MPC is removed and the trained ANN successfully regulates the voltage of the DC/DC converter as per reference voltage. The bayesian regularized technique is used to train the ANN and adjust the biases and weights of the ANN. Different types of tests were also performed during simulation, such as step change of load, the shift of load from full load to no load, and vice versa, in order to demonstrate the performance of the proposed controller. It has been observed through simulation results that the overall performance of the proposed control scheme is better than the classical linear controllers. The implementation of the proposed technique would be useful in DC microgrid applications, where the DC boost converters require high accuracy for tuning controller parameters.

\section{Acknowledgment}
This work is carried out by the financial support provided by the Walter Ahlström Foundation Finland with grant No. $2021/40$. Some parts of this work are done in the SolarX research project with the financial support provided by Business Finland with Grant No. $6844/31/2018$. The financial support provided through these funding organizations is highly acknowledged.
\bibliographystyle{IEEEtran}
\bibliography{References}
\end{document}